\documentclass{article}%
\usepackage{amsmath}
\usepackage{amsfonts}
\usepackage{amssymb}
\usepackage{graphicx}%
\begin{document}

\title{The Meservey-Tedrov effect in FSF double tunneling junctions}
\date{\today}
\author{Gerd Bergmann$^{1}$, Jia Lu$^{2}$ and Dawei Wang$^{2}$\\$^{1}$ Physics Department, Univ.South.California\\Los Angeles, CA 90089-0484, USA\\$^{2}$Department of Materials Science and of \\Electrical Engineering, University of Californie, \\Irvine, CA 92697, USA}
\maketitle

\begin{abstract}
Double tunneling junctions of ferromagnet-superconductor-ferromagnet
electrodes\ (FSF) show a jump in the conductance when a parallel magnetic
field reverses the magnetization of one of the ferromagnetic electrodes. This
change is generally attributed to the spin-valve effect or to pair breaking in
the superconductor because of spin accumulation. In this paper it is shown
that the Meservey-Tedrov effect causes a similar change in the conductance
since the magnetic field changes the energy spectrum of the quasi-particles in
the superconductor. A reversal of the bias reverses the sign in the
conductance jump.

PACS: 73.50.-h, 73.50.Bk, 73.23.-b, 73.25.+i

\end{abstract}

\section{Introduction}

During the last five years single electron transistors (SET) with two
ferromagnetic electrodes and a superconducting island have been studied
experimentally \cite{C20}, \cite{H30}, \cite{H29}, \cite{J10} and
theoretically \cite{M52}. Experimentally one generally uses
ferromagnetic-superconductor-ferromagnetic double junctions which consist of
an Al strip of length of about 1$\mu m$, width of 50-100$nm$ and thickness of
about 20$nm$. The Al is oxidized, and two Co electrodes with slightly
different width and twice the thickness cross the Al strip at a separation of
a few 100$nm$. They form two tunneling junctions. Fig.1 shows the schematic
arrangement of the two ferromagnetic Co electrodes and Al island. A magnetic
field is applied parallel to the Co strips and aligns the magnetization of the
two Co strips. Then the magnetic field is reversed. At a magnetic field
$B_{sw}$ the wider Co strip flips its magnetization to be parallel to the
magnetic field while the narrower Co strip remains\ anti-parallel to the
external field because its coercitive field is larger. At the same time the
current through the double junction changes abruptly at $B_{sw}$. At the field
$B_{sn}$ the narrower Co strip also reverses its magnetization and the
magnetizations of the two Co strips are again parallel to the external field.
(The relative orientations of the magnetic field and the magnetization of the
two Co electrodes is shown in Fig.4). If one applies constant bias to the
junction then the current shows a jump at each of the fields $B_{sw}$ and
$B_{sn}$ (with opposite sign). Such jumps at the fields $B_{sw},B_{sn}$ have
been observed in a number of experiments \cite{C20}, \cite{H30}, \cite{H29},
\cite{L32}.

In the theoretical discussion one generally considers two mechanisms which
change the current (i.e. conductance) of the double junction in the field
range $\left(  B_{sw},B_{sn}\right)  $:

\begin{enumerate}
\item Spin-valve effect: When the magnetizations of the Co strips
$\mathbf{m}_{1}$ and $\mathbf{m}_{2}$ are both parallel to $\widehat
{\mathbf{y}}$ then one has a large density of states in both Co electrodes for
the spin moment up electrons, while the spin moment down electrons have a
small density of states in both electrodes. For the (spin) moment up one has
two small resistances $R_{t\uparrow}$ in series and for the other direction
two large resistances $R_{t\downarrow}$. The total conductance is then
$G_{\upharpoonleft\text{ }\upharpoonright}=1/\left(  2R_{t\uparrow}\right)
+1/\left(  2R_{t\downarrow}\right)  $. If the two Co strips have opposite
magnetization then the conductance is $G_{\upharpoonleft\text{ }%
\downharpoonright}=2/\left(  R_{t\uparrow}+R_{t\downarrow}\right)  $. It is
easy to show that $G_{\upharpoonleft\text{ }\upharpoonright}\geq
G_{\upharpoonleft\text{ }\downharpoonright}$. Therefore the current should
drop inside the field window \cite{M51}, \cite{M53}.

\item Gap reduction due to spin accumulation: In the anti-ferromagnetic
alignment one obtains spin moment accumulation \cite{M52} because the spin
moment up electrons have a small resistance for tunneling onto the island and
a large resistance to tunnel off the island while the opposite is true for
spin moment down electrons. This spin moment accumulation can reduce the
superconducting gap of the Al island. This will lead to an increase of the
conductance in the field window $\left(  B_{sw},B_{sn}\right)  $.
\end{enumerate}

\[%
\begin{tabular}
[c]{l}%
{\includegraphics[
height=2.3993in,
width=2.313in
]%
{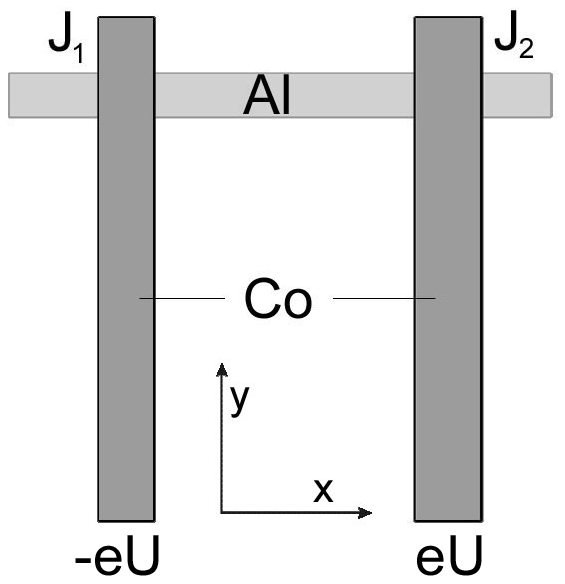}%
}%
\\
Fig.1: The schematic geometry of an FSF double junction,\\
consisting of Co/Al/Co.
\end{tabular}
\ \ \ \ \ \ \
\]

In this paper we want to show theoretically that there is an additional
contribution to the current because of the Zeeman effect which shifts the
excitation spectrum of the quasi particles in the Al by $\overrightarrow
{\mu_{e}}\mathbf{B}$ ($\overrightarrow{\mu_{e}}$=moment of the spin up and
down electrons, $\mathbf{B}$=external magnetic field). This effect has been
intensively studied by Meservey and Tedrow in many beautiful experiments (see
the review article \cite{M34}). In a series of papers \cite{M55}, \cite{M56},
\cite{M57} their group investigated the tunneling I-V-curve for
ferromagnet-superconductor tunneling junctions in different magnetic field.
They showed that the I-V-curves were asymmetric with respect to the voltage
(because of the different density of spin up and down electron at the Fermi
surface). From the asymmetry they derived the polarization of the effective
density of states of the tunneling electrons. To our knowledge the magnetic
field and the magnetization were always parallel to each other in their measurements.

In this paper we want to demonstrate that one has to take the field and spin
dependence of the quasi-particles in the Al into account when calculating the
current through the double tunneling junction. To demonstrate this effect we
consider a double tunneling junction in which the two junctions are so far
separated that the spin-orbit scattering destroys any spin polarization along
the diffusion path of the electrons from the first tunneling junction to the
second one. This means that only the total current $I_{1}$ through junction
one must be equal to the total current $I_{2}$ through junction two; the spin
up and down currents through the two junctions can be quite different.

\section{Theory and Simulation}

\subsection{Single junctions}

We first consider a single ferromagnet-superconductor tunneling junction at
zero temperature. In Fig.2 the density of states for both metals is plotted
after lifting the energy bands of the ferromagnet by $eU$.
\[%
\begin{tabular}
[c]{l}%
{\includegraphics[
height=2.7239in,
width=4.4774in
]%
{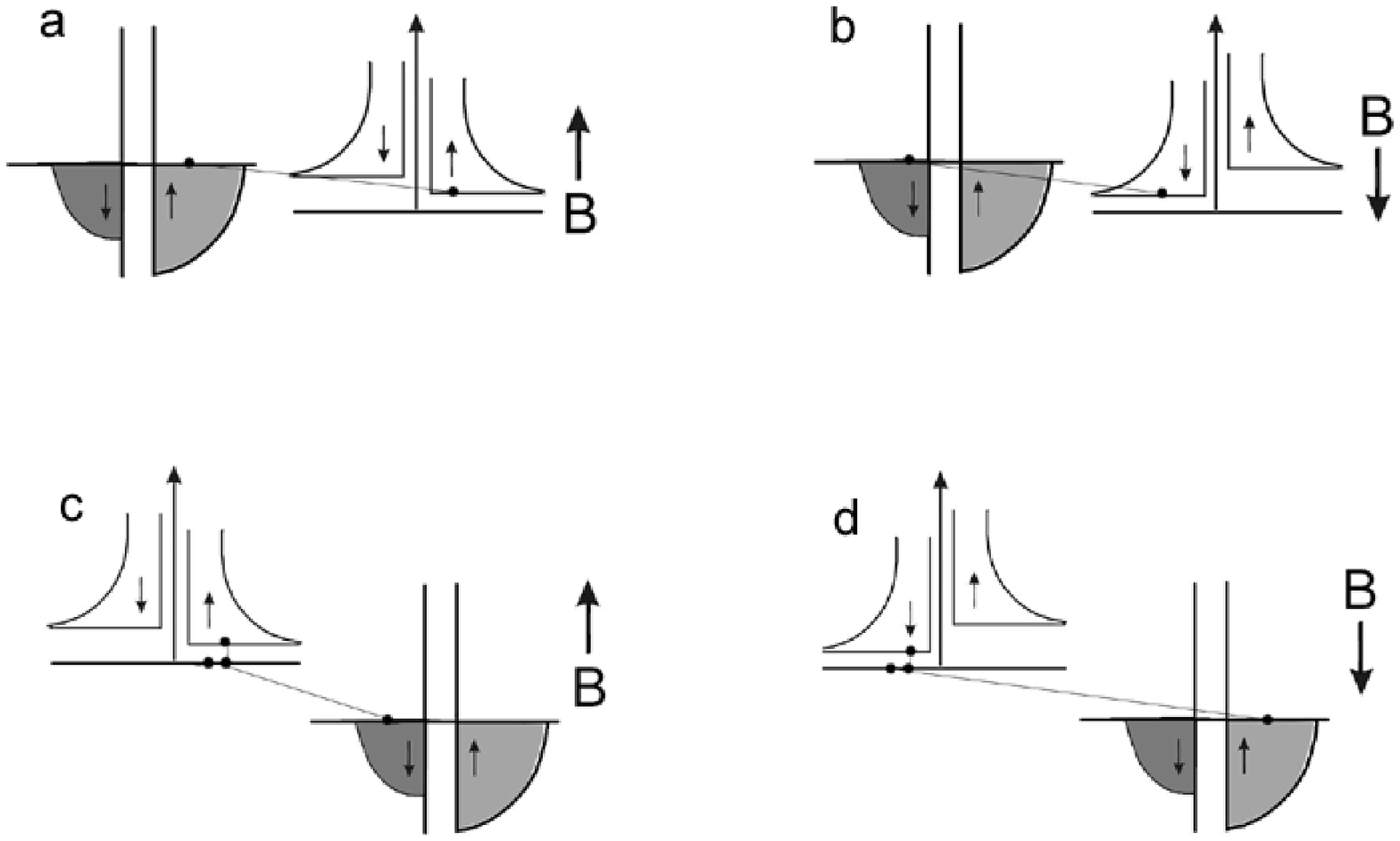}%
}%
\\
Fig.2: The tunneling density of spin moment up electrons in\\
an FS tunneling junction for different orientations of the\\
magnetic field and the magnetization $\mathbf{m}$ of the ferromagnet.\\
a) $\mathbf{B}$ and $\mathbf{m}$ are parallel, both pointing in $\widehat
{\mathbf{y}}$,\\
b) $\mathbf{B}$ and $\mathbf{m}$ are anti-parallel, $\mathbf{B}$ pointing in
$-\widehat{\mathbf{y}}$, $\mathbf{m}$ in $\widehat{\mathbf{y}}$\\
c,d) The bias is reversed.
\end{tabular}
\ \ \ \ \ \ \ \
\]

In large body of experiments Merservey and Tedrow \cite{M34} showed that a
magnetic field parallel to the tunneling junction shifts the excitation
spectra of spin up and down electrons in the superconductor by
$\overrightarrow{\mu_{e}}\mathbf{B}$ in opposite directions. This enhances the
current of the majority spin (see Fig.2a) when the electrons are flowing from
the ferromagnet to the superconductor. One obtains a spin current (with moment
up). The I-V-curve is not (point-) symmetric about the origin.

The tunneling current is for spin \emph{moment} up and down is given at zero
temperature by%
\begin{align*}
I_{\uparrow}  &  =CN_{M}N_{S}\int_{\Delta-\overrightarrow{\mu_{e}}\mathbf{B}%
}^{eU}\frac{\left(  E+\overrightarrow{\mu_{e}}\mathbf{B}\right)  }%
{\sqrt{\left(  E+\overrightarrow{\mu_{e}}\mathbf{B}\right)  ^{2}-\Delta\left(
B\right)  ^{2}}}dE=CN_{M}N_{S}\sqrt{\left(  \left(  eU+\mu_{B}B\right)
\right)  ^{2}-\Delta\left(  B\right)  ^{2}}\\
I_{\downarrow}  &  =CN_{m}N_{S}\int_{\Delta-\overrightarrow{\mu_{e}}%
\mathbf{B}}^{eU}\frac{\left(  E+\overrightarrow{\mu_{e}}\mathbf{B}\right)
}{\sqrt{\left(  E+\overrightarrow{\mu_{e}}\mathbf{B}\right)  ^{2}%
-\Delta\left(  B\right)  ^{2}}}dE=CN_{m}N_{S}\sqrt{\left(  \left(  eU-\mu
_{B}B\right)  \right)  ^{2}-\Delta\left(  B\right)  ^{2}}%
\end{align*}

Here $N_{M}$ and $N_{m}$ are the majority and minority density of states in
the ferromagnet and $N_{S}$ is the density of states of the superconductor in
the normal state. In the superconducting state in the presence of a magnetic
field $N_{S}$ is modified by the factor $\left(  E+\overrightarrow{\mu_{e}%
}\mathbf{B}\right)  /\sqrt{\left(  E+\overrightarrow{\mu_{e}}\mathbf{B}%
\right)  ^{2}-\Delta\left(  B\right)  ^{2}}$. The constant $C$ contains the
tunneling matrix elements and universal constants. The energy gap is given by
$\Delta\left(  B\right)  .$ For thin films and stripes which are aligned
parallel to an external magnetic field we use the result from reference
\cite{D39} for the dependence of $\Delta$ on the magnetic field :%
\[
\Delta\left(  T,B\right)  =\Delta\left(  T,0\right)  \sqrt{1-\left(  \frac
{B}{B_{c}\left(  T\right)  }\right)  ^{2}}%
\]
where the field $B_{c}\left(  T\right)  $ is determined by the ratio of the
penetration depth $\lambda\left(  T\right)  $ and the film thickness.
\[
B_{c}\left(  T\right)  =\sqrt{24}\frac{\lambda\left(  T\right)  }{d}%
B_{cb}\left(  T\right)
\]
$B_{cb}\left(  T\right)  $ is the thermodynamic critical field.

The use of the density of states in the tunneling current is a dramatic
oversimplification since the tunneling probability of electrons at different
parts of the Fermi surface depends strongly on the direction of their group
velocity relative to the tunneling barrier. So $N_{M}$, $N_{m}$ and $N_{S}$
have to be interpreted as "effective tunneling densities of states". In the
present paper we only need the relative magnitudes of $N_{M}$ and $N_{m}$
which are given by the experimental polarization of the tunneling current.

Merservey and Tedrow obtained a number of interesting results for an FS
junction in a parallel magnetic field:

\begin{itemize}
\item The I-V-tunneling curve is not (point) symmetric about the origin.

\item The tunneling current is polarized and the polarization can be evaluated.

\item The polarization is always parallel to the majority moment of the
ferromagnet and not proportional to the d-density of states at the Fermi surface.
\end{itemize}

They obtained a polarization of 35\% for Co/Al junctions.

There is another interesting consequence of the energy shift of the Zeeman
effect. Let us consider a single Co/Al tunneling junction, i.e. the left half
of the Fig.1. We align the magnetic field parallel to the Co strip in the
negative y-direction and keep the voltage across the junction constant. For
simplicity we assume that the temperature is (close to) zero. We start with
the magnetic field $-B_{c}\left(  0\right)  $ which suppresses
superconductivity in the Al completely. Then we sweep the magnetic field
towards $+B_{c}\left(  0\right)  $. As soon as the magnetic field takes
positive values the magnetization of the Co and the field are anti-parallel
and therefore the junction is in an instable energetic state. Due to its
coercitive field $B_{sw}$ the Co can maintain the anti-parallel orientation up
to the field $B_{sw}$. Then the Co film will switch its magnetization. As a
consequence the tunneling current will also change.

In Fig.3 we calculate the current through the junction using the following
parameters: the energy gap in Al at zero temperature $\Delta=0.2meV$, the
field that suppresses superconductivity completely $B_{c}=1.5T$, the switching
field $B_{sw}=0.16T$, the polarization of the effective density in the Co
$p=.35$. We sweep the external magnetic field from $-0.5T$ \ to $0.5T$. When
the magnetic field changes sign the Zeeman term changes sign as well. At the
magnetic field $B_{sw}$ the direction of the Co magnetization $\mathbf{m}$
becomes aligned parallel to the magnetic field. At the same time the current
jumps to a higher value.
\[%
\begin{tabular}
[c]{l}%
\raisebox{-0pt}{\includegraphics[
height=3.0369in,
width=4.3404in
]%
{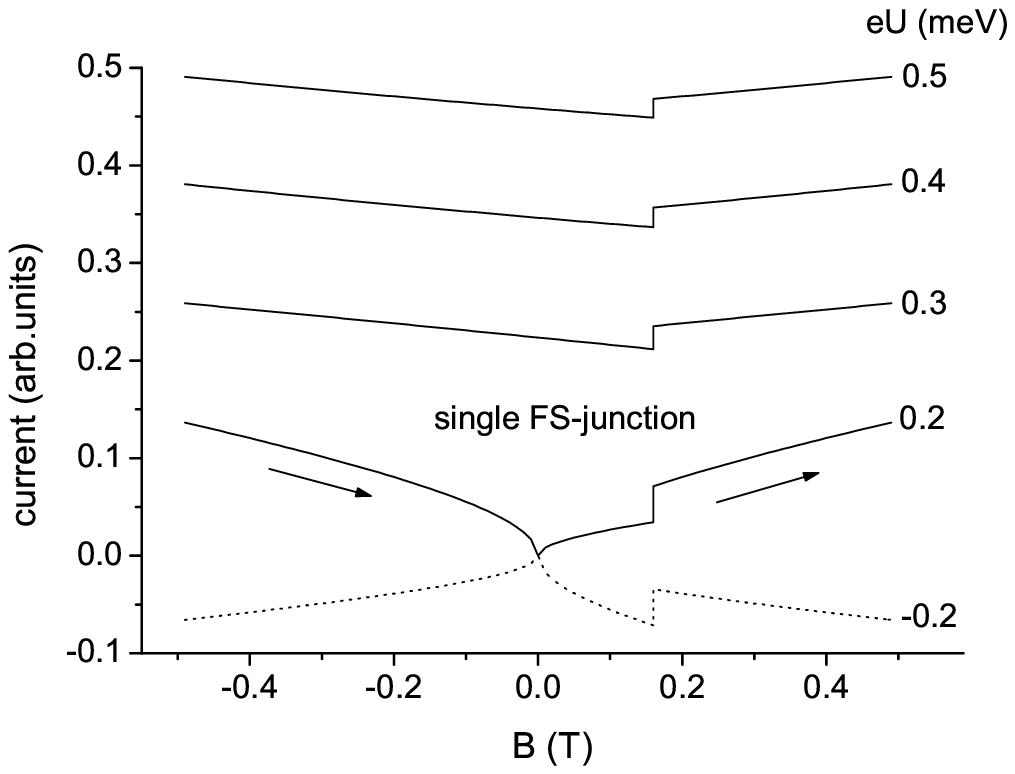}%
}%
\\
Fig.3: The simulated current through an FS-tunneling junction\\
while the magnetic field sweeps from $-0.5T$ to $+0.5T$.\\
The numbers at the curves give the different bias. At\\
$B=0.16T$ the magnetization of the Co strip reverses.
\end{tabular}
\ \ \ \ \ \ \ \ \ \
\]

In Fig.3 the calculated tunneling current through a Co/Al junction is plotted
for constant bias as a function of the magnetic field. The different curves
are for different bias which is given in $meV$ at the right side of the
curves. One recognizes that the current shows a jump at the switching field
$B_{sw}=0.16T$. For positive bias the current increases at the switching field
while for negative bias the (absolute value of the) current decreases.
Furthermore the minimum of the the I-B-curve is not at $B=0$ but shifted to
positive field values. If the field is then swept from $B_{m}=0.5T$ to $-0.5T$
the resulting current curves are just a mirror image of the shown curves.

It is important to note that a reversal of the applied voltage corresponds to
a tunneling of electrons from the superconductor to the ferro-magnet (see
Fig.2b,c). In this case the current is smaller if $\mathbf{m}$ is parallel to
$\mathbf{B}$ because for an electron with moment up to tunnel from S to F a
Cooper pair has to split and the moment down electron is elevated by
$\Delta+\mu_{B}B$ into an excited state in the superconductor while the moment
up electron tunnels into the ferromagnet. The contribution of moment up
electrons to the tunneling current is reduced to $CN_{M}N_{S}\sqrt{\left(
\left(  eU-\mu_{B}B\right)  \right)  ^{2}-\Delta\left(  B\right)  ^{2}}$.
Therefore the current changes to a smaller value when the magnetization flips.

\subsection{Double junctions}

We now consider the double junction in Fig.1. However, we use a long Al strip
so that the two junctions are relatively far apart. As we discussed above the
tunneling current through a single junction Co/Al is polarized. This means
that one has a source of polarized electrons in the Al strip. Meakawa et al.
\cite{M53} calculated an (opposite) shift in the chemical potential for (spin)
moments up and down. The lifetime of an electron in a given spin state is
limited by the spin-orbit scattering and the spin state decays as $\exp\left(
-t/\tau_{sf}\right)  $ where $\tau_{sf}$ is the spin-flip lifetime which is
3/2 times the spin-orbit scattering time $\tau_{so}$. The injected spin
polarization at the left tunneling junction diffuses through the Al strip
towards the right junction and vice versa. Along this diffusion path the spin
decays. For the resulting spin diffusion length one finds different orders of
magnitude in the literature, varying between 10-100$nm$ and 1$\mu m$
\cite{M34}, \cite{J7}.

As shown in Fig.1 the total potential difference between the right and the
left electrode is $2eU.$ We consider the bias as positive when the potential
on the right electrode is positive. Then the electrons flow from the left to
the right side as shown in Fig.4.%
\[%
{\includegraphics[
height=4.7696in,
width=3.4769in
]%
{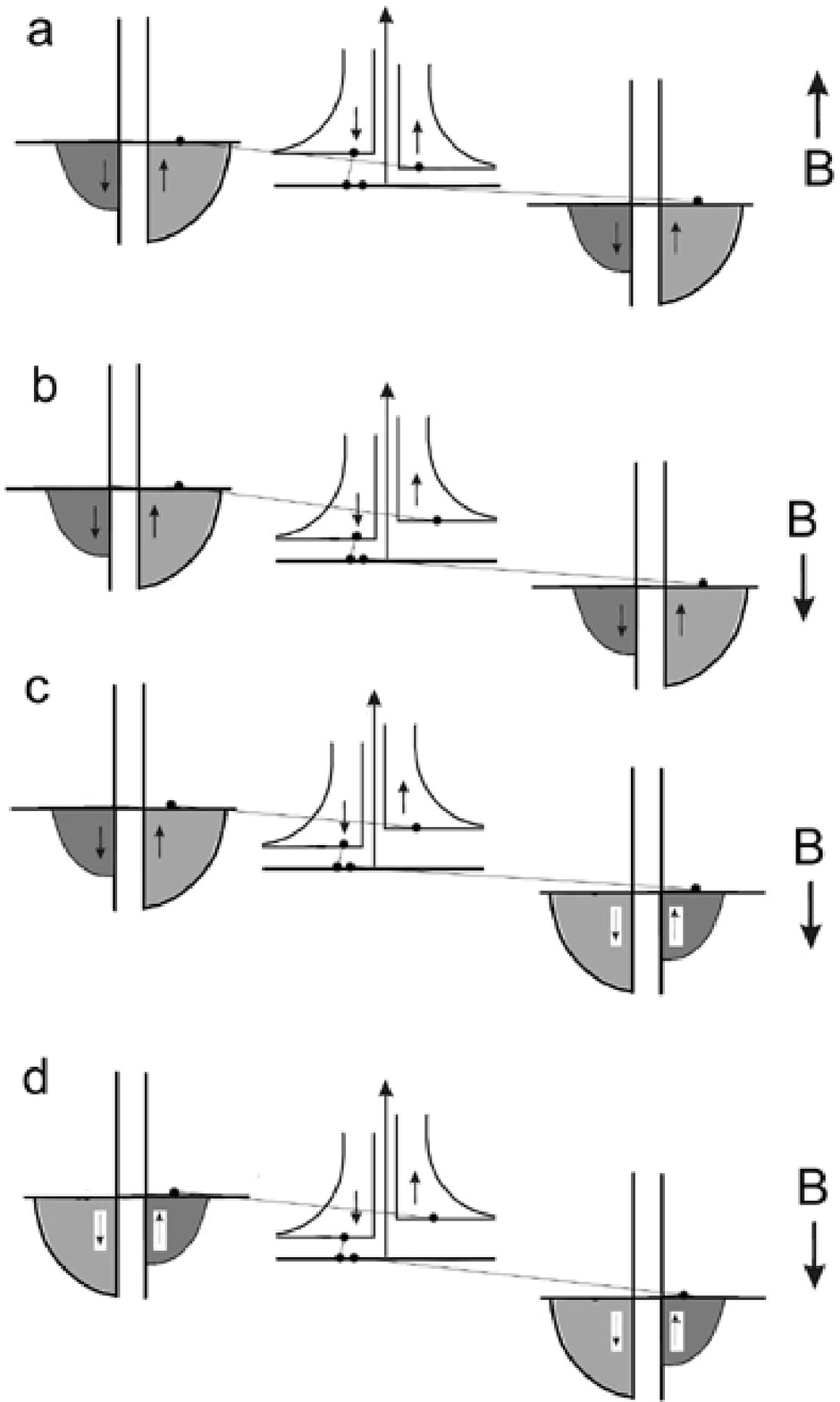}%
}%
\ \ \
\]%
\[%
\begin{tabular}
[c]{l}%
Fig.4: The current of spin moment up electrons through an FSF-double
junction.\\
a) The moments $\mathbf{m}_{1}$, $\mathbf{m}_{2}$ and $\mathbf{B}$ are all
parallel, pointing in $+\widehat{\mathbf{y}}$ direction,\\
b) the magnetic field has changed to the $-\widehat{\mathbf{y}}$ direction,\\
c) the moment $\mathbf{m}_{2}$ has switched at $B_{sw}$ to the $-\widehat
{\mathbf{y}}$ direction,\\
d) the moment $\mathbf{m}_{1}$ has switched at $B_{sn}$ to the $-\widehat
{\mathbf{y}}$ direction,\\
$\mathbf{m}_{1}$, $\mathbf{m}_{2}$ and $\mathbf{B}$ are all parallel, pointing
in $-\widehat{\mathbf{y}}$ direction
\end{tabular}
\ \ \ \ \
\]

In contrast to the goal of the spin valve we choose a large separation of the
two junctions (in our virtual experiment) so that each junction is unaffected
by the polarized current of the other. In this case we have excluded all
spin-valve effects in our virtual experiment. Mathematically this requirement
is expressed by the condition that only the total currents through junctions
J$_{1}$ and J$_{2}$ must be identical; the individual spin currents can be
different. As a result both spin directions experience the same shift in the
chemical potential.

We consider first the special case (a) in Fig.4 where $\mathbf{m}_{1}$,
$\mathbf{m}_{2}$ and $\mathbf{B}$ are all pointing in the positive
$\widehat{\mathbf{y}}$-direction. In general the currents through the
junctions J$_{1}$ and J$_{2}$ are not identical when their bias is the same,
i.e. $eU$. Therefore the chemical potential of the island will shift by $\phi$
(which has to be determined self consistently). Then the (spin) moment up
current through junctions J$_{1}$ and J$_{2}$ are given by%
\begin{align}
I_{1\uparrow}  &  =CN_{S}N_{m}\sqrt{\left(  \left(  V_{e}+\phi+\mu
_{B}B\right)  ^{2}-\Delta^{2}\right)  }\label{SpCrt1}\\
I_{2\uparrow}  &  =CN_{S}N_{m}\sqrt{\left(  \left(  V_{e}-\phi-\mu
_{B}B\right)  ^{2}-\Delta^{2}\right)  }\nonumber
\end{align}
(The symbol $\uparrow$ stands again for spin moment up.)

The other current contributions can be obtained from these currents by
applying simple rules:

\begin{enumerate}
\item The contribution of spin moment down electrons is obtained changing the
sign of the term $\mu_{B}B$ in $I_{1}$ and $I_{2}$ and exchanging $N_{M}$ and
$N_{m}$.

\item If $\mathbf{m}_{1}$ points in the $-\widehat{\mathbf{y}}$-direction one
has to replace $N_{M}$ by $N_{m}$ in $I_{1}$.

\item If $\mathbf{m}_{2}$ points in the $-\widehat{\mathbf{y}}$-direction one
has to replace $N_{M}$ by $N_{m}$ in $I_{2}$.

\item If $\mathbf{B}$ points in the $-\widehat{\mathbf{y}}$-direction one has
to change the sign of the term $\mu_{B}B$ in $I_{1}$ and $I_{2}$.
\end{enumerate}

It is sufficient to calculate the current for spin moment up (equation
$\left(  \ref{SpCrt1}\right)  $) in the alignment of Fig.4a. Then the above
rules yield the current for moment up and down under all circumstances. For
example the corresponding spin moment down currents are%
\begin{align}
I_{1\downarrow}  &  =CN_{S}N_{M}\sqrt{\left(  \left(  V_{e}+\phi-\mu
_{B}B\right)  ^{2}-\Delta^{2}\right)  }\label{SpCrt2}\\
I_{2\downarrow}  &  =CN_{S}N_{M}\sqrt{\left(  \left(  V_{e}-\phi+\mu
_{B}B\right)  ^{2}-\Delta^{2}\right)  }\nonumber
\end{align}

We calculate the total current perturbatively. For sufficiently large bias,
i.e., $eU>\Delta$ the terms $e\phi$ and $\mu_{B}B$ are small compared to
$\sqrt{V_{e}^{2}-\Delta^{2}}$ and we can expand the different current
contributions as a Taylor series in terms of $e\phi$ and $\mu_{B}B$ up to
second order. Since the current depends on the orientation of three vectors,
$\mathbf{m}_{1},\mathbf{m}_{2}$ and $\mathbf{B}$ we choose the $\widehat
{\mathbf{y}}$-direction as a reference direction. The value of $B$ is negative
when $\mathbf{B}$ is anti-parallel to $\widehat{\mathbf{y}}$. We calculate (in
perturbation)\ the currents $I_{\upharpoonleft\text{ }\upharpoonright
},I_{\upharpoonleft\text{ }\downharpoonright},I_{\downharpoonleft\text{
}\upharpoonright},I_{\downharpoonleft\text{ }\downharpoonright\text{ }}$ for
the four possible orientations of $\mathbf{m}_{1}$ and $\mathbf{m}_{2}$. The
results are collected in the following equations
\begin{align}
I_{\upharpoonleft\text{ }\upharpoonright}  &  =CN_{S}\left(  N_{M}%
+N_{m}\right)  \sqrt{V_{e}^{2}-\Delta^{2}}\left(  1-\frac{2N_{M}N_{m}}{\left(
N_{M}+N_{m}\right)  ^{2}}\frac{\Delta^{2}}{\left(  V_{e}^{2}-\Delta
^{2}\right)  ^{2}}\allowbreak\left(  \mu_{B}B\right)  ^{2}\right)
\label{AllCurrnt}\\
I_{\upharpoonleft\text{ }\downharpoonright}  &  =CN_{S}\left(  N_{M}%
+N_{m}\right)  \sqrt{V_{e}^{2}-\Delta^{2}}\left(  1+\frac{\left(  N_{M}%
-N_{m}\right)  }{\left(  N_{M}+N_{m}\right)  }\frac{V_{e}}{\left(  V_{e}%
^{2}-\Delta^{2}\right)  }\mu_{B}B\allowbreak-\frac{1}{2}\allowbreak
\frac{\Delta^{2}}{\left(  V_{e}^{2}-\Delta^{2}\right)  ^{2}}\left(  \mu
_{B}B\right)  ^{2}\right) \nonumber\\
I_{\downharpoonleft\text{ }\upharpoonright}  &  =CN_{S}\left(  N_{M}%
+N_{m}\right)  \sqrt{V_{e}^{2}-\Delta^{2}}\left(  1-\frac{\left(  N_{M}%
-N_{m}\right)  }{\left(  N_{M}+N_{m}\right)  }\frac{V_{e}}{\left(  V_{e}%
^{2}-\Delta^{2}\right)  }\mu_{B}B-\frac{1}{2}\allowbreak\frac{\Delta^{2}%
}{\left(  V_{e}^{2}-\Delta^{2}\right)  ^{2}}\left(  \mu_{B}B\right)
^{2}\right) \nonumber\\
I_{\downharpoonleft\text{ }\downharpoonright}  &  =CN_{S}\left(  N_{M}%
+N_{m}\right)  \sqrt{V_{e}^{2}-\Delta^{2}}\left(  1-2\frac{N_{M}N_{m}}{\left(
N_{M}+N_{m}\right)  ^{2}}\frac{\Delta^{2}}{\left(  V_{e}^{2}-\Delta
^{2}\right)  ^{2}}\left(  \mu_{B}B\right)  \allowbreak^{2}\right) \nonumber
\end{align}
Here the indices of the currents give the direction of the magnetizations
$\mathbf{m}_{1}$ and $\mathbf{m}_{2}$ with respect to the $\widehat
{\mathbf{y}}$ direction, for example $I_{\downharpoonleft\text{ }%
\upharpoonright}$ is the current for $\mathbf{m}_{1}$ anti-parallel and
$\mathbf{m}_{2}$ parallel to $\widehat{\mathbf{y}}$. When $\mathbf{m}_{1}$ and
$\mathbf{m}_{2}$ are anti-parallel to each other (i.e. for $I_{\upharpoonleft
\text{ }\downharpoonright}$ and $I_{\downharpoonleft\text{ }\upharpoonright}$)
then the chemical potential of the island $e\phi$ is zero. In the parallel
orientation one obtains
\[%
\begin{tabular}
[c]{lll}%
$%
\begin{array}
[c]{c}%
e\phi=-\mu_{B}B\frac{\left(  N_{M}-N_{m}\right)  }{\left(  N_{M}+N_{m}\right)
}%
\end{array}
$ & for & $I_{\upharpoonleft\text{ }\upharpoonright}$\\
&  & \\
$e\phi=+\mu_{B}B\frac{\left(  N_{M}-N_{m}\right)  }{\left(  N_{M}%
+N_{m}\right)  }$ & for & $I_{\downharpoonleft\text{ }\downharpoonright}$%
\end{tabular}
\ \
\]

The dependence of the currents on the quadratic term $\left(  \mu_{B}B\right)
^{2}$ is rather weak, so that it is sufficient for a qualitative discussion to
restrict ourselves to the linear dependence on $\left(  \mu_{B}B\right)  $.
When we start with a large negative magnetic field $\left(  \mathbf{B}%
\upharpoonleft\downharpoonright\widehat{\mathbf{y}}\right)  $ then both
magnetizations, $\mathbf{m}_{1}$ and $\mathbf{m}_{2}$, are anti-parallel to
$\widehat{\mathbf{y}}$. In the linear approximation the current is
$I_{\downharpoonleft\text{ }\downharpoonright}\thickapprox\sqrt{V_{e}%
^{2}-\Delta^{2}}\left(  N_{M}+N_{m}\right)  $. At the positive field $B_{sw}$
the magnetization of the junction J$_{\text{2}}$ flips and aligns parallel to
the field. Then the new current is $I_{\downharpoonleft\text{ }\upharpoonright
}$ which corresponds to a relative decrease of the current
\[
\frac{\Delta I}{I}=-\frac{\left(  N_{M}-N_{m}\right)  }{\left(  N_{M}%
+N_{m}\right)  }\frac{V_{e}}{\left(  V_{e}^{2}-\Delta^{2}\right)  }\mu
_{B}B\allowbreak
\]
At the higher field $B_{sn}$ the other electrode J$_{\text{1}}$ also aligns
parallel to $\widehat{\mathbf{y}}.$ Then the current returns to the original
curve since $I_{\upharpoonleft\text{ }\upharpoonright}=I_{\downharpoonleft
\text{ }\downharpoonright}$ in this approximation. It is important to point
out that the jump in the current is positive when the magnetization of
junction J$_{\text{1}}$ (i.e. $\mathbf{m}_{1}$) flips first. Then the current
changes from $I_{\downharpoonleft\text{ }\downharpoonright\text{ }}$ to
$I_{\upharpoonleft\text{ }\downharpoonright}$.

In Fig.5 we sweep the magnetic field from $-0.5T$ to $+0.5T$. The current
through the double junction is plotted versus the sweeping magnetic field. The
junction J$_{2}$ has the switching field $B_{sw}=0.14T$ while junction J$_{1}$
has the larger switching field of $B_{sn}=0.18T$. We use different bias
voltages in the range of $-0.2meV\leq-eU\leq0.7meV$. The $I-B$-curve in Fig.5
shows a downward displacement in the field window $\left(  B_{sw}%
,B_{sn}\right)  $. For negative $-eU$ \ (i.e. $U>0$) the absolute value of the
current increases. This means that the displacement changes sign when the bias
is reversed.%

\[%
\begin{tabular}
[c]{l}%
{\includegraphics[
height=2.9398in,
width=3.8223in
]%
{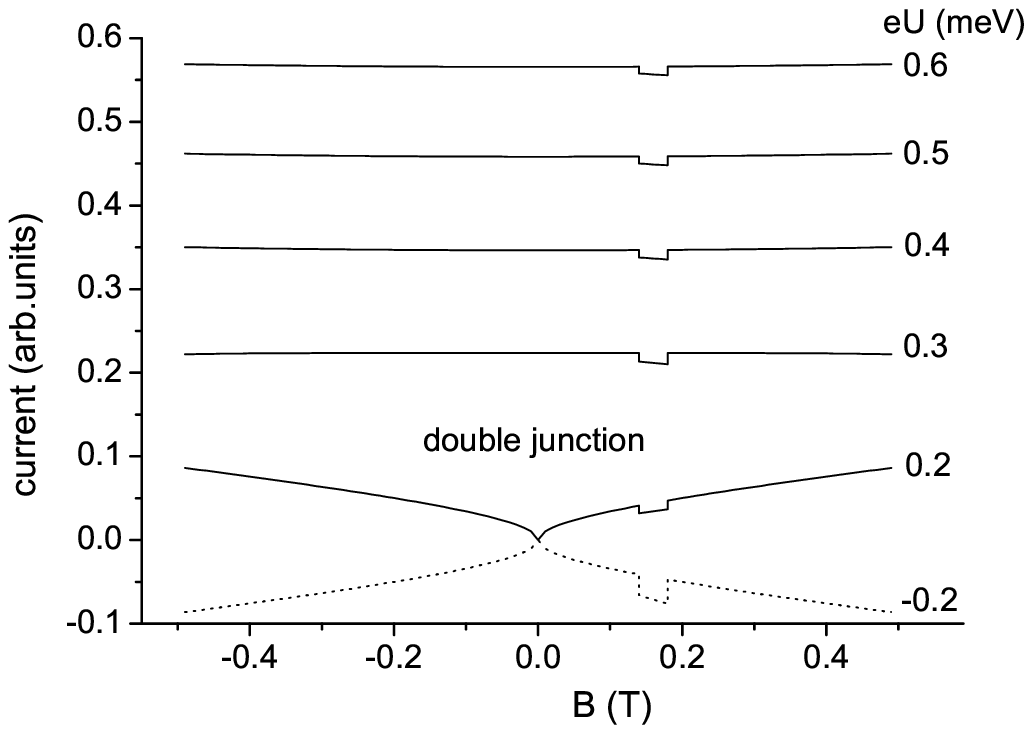}%
}%
\\
Fig.5: The current through an FSF-double junction while the magnetic field\\
sweeps from $-0.5T$ to $0.5T$. The numbers at the curves give the different\\
bias. The switching fields for the two Co strips are $0.14T$ and $0.18T$.
\end{tabular}
\ \ \ \ \ \ \ \ \ \ \
\]

\section{The single electron transistor}

When the size of the two tunneling junctions is in the nano-meter scale then
their capacity is small and the tunneling electrons change the Coulomb energy
on the island \cite{L30}, \cite{N12}, \cite{G40}, \cite{S46}, \cite{S47},
\cite{K50}. If an electron from the left electrode with the band energy
$\varepsilon_{L}$ tunnels onto a state on the island with the band energy
$\varepsilon_{I}$ and changes the number of electrons on the island from $n$
to $\left(  n+1\right)  $ then conservation of energy requires that
\[
\varepsilon_{L}+Ue=\varepsilon_{I}+\left(  2n+1\right)  \frac{e^{2}%
}{2C_{\Sigma}}-\frac{C_{G}}{C_{\Sigma}}eU_{G}%
\]
On the other hand if an electron from the island with the band energy
$\varepsilon_{I}$ tunnels into a state on the right electrode with the band
energy $\varepsilon_{R}$ and changes the number of electrons from $n$ to
$\left(  n-1\right)  $ one has to fulfill the condition%
\[
\varepsilon_{I}+Ue=\varepsilon_{R}-\left(  2n-1\right)  \frac{e^{2}%
}{2C_{\Sigma}}-\frac{C_{G}}{C_{\Sigma}}eU_{G}%
\]
Here $U_{G}$ is the gate voltage and $C_{\Sigma}^{-1}=C_{1}^{-1}+C_{2}%
^{-1}+C_{G}^{-1}$ where $C_{1}$, $C_{2}$ and $C_{G}$ are the capacities of the
two tunneling junctions and the gate. The Coulomb blockade energy is given by
$E_{Cb}=e^{2}/2C_{\Sigma}$. In the following we consider only zero gate voltage.

At a sufficiently large bias ($2eU>2\left(  \Delta+E_{Cb}\right)  $) the
island can gain or loose up to $n_{0}$ electrons where (at $T=0)$ $n_{0}$ is
given by $n_{0}=$int$\left[  \frac{1}{2}\left(  \left(  eU-\Delta\right)
/E_{Cb}-1\right)  \right]  $. The probability for $n$ excess electrons on the
island may be $p\left(  n\right)  $ which will be determined self-consistently.

First we calculate the currents for spin moment up and $\mathbf{m}_{1}$ and
$\mathbf{m}_{2}$ parallel to $\widehat{\mathbf{y}}$. For tunneling from the
left Co electrode onto the Al island with $n$ excess electrons (prior to the
tunneling) the current $I_{1\uparrow}\left(  n\right)  $ of moment up
electrons is
\[
I_{1\uparrow}\left(  n\right)  =p\left(  n\right)  CN_{M}N_{S}\sqrt{\left(
\left(  eU-\left(  2n+1\right)  E_{Cb}+\mu_{B}B\right)  \right)  ^{2}%
-\Delta^{2}}%
\]
The current from the Al island with $n$ excess electrons onto the right Co
electrode is
\[
I_{2\uparrow}\left(  n\right)  =p\left(  n\right)  CN_{M}N_{S}\sqrt{\left(
\left(  eU+\left(  2n-1\right)  E_{Cb}-\mu_{B}B\right)  \right)  ^{2}%
-\Delta^{2}}%
\]
The occupation probabilities $p\left(  n\right)  $ are obtained by the
condition that the flow of electrons on the island with $n$ excess electrons
is equal to the out-flow (see for example the review article \cite{S46}). This
yields simple linear equations for $p\left(  n\right)  $. The currents for
spin moment down and different orientations of $\mathbf{B}$, $\mathbf{m}_{1}$
and $\mathbf{m}_{2}$ are obtained by applying the rules which we stated above.
The results of this calculation are plotted in Fig.6. We use for the Coulomb
energy $E_{Cb}$ the value $E_{Cb}=0.101meV$.
\begin{align*}
&
{\includegraphics[
height=3.1756in,
width=4.1062in
]%
{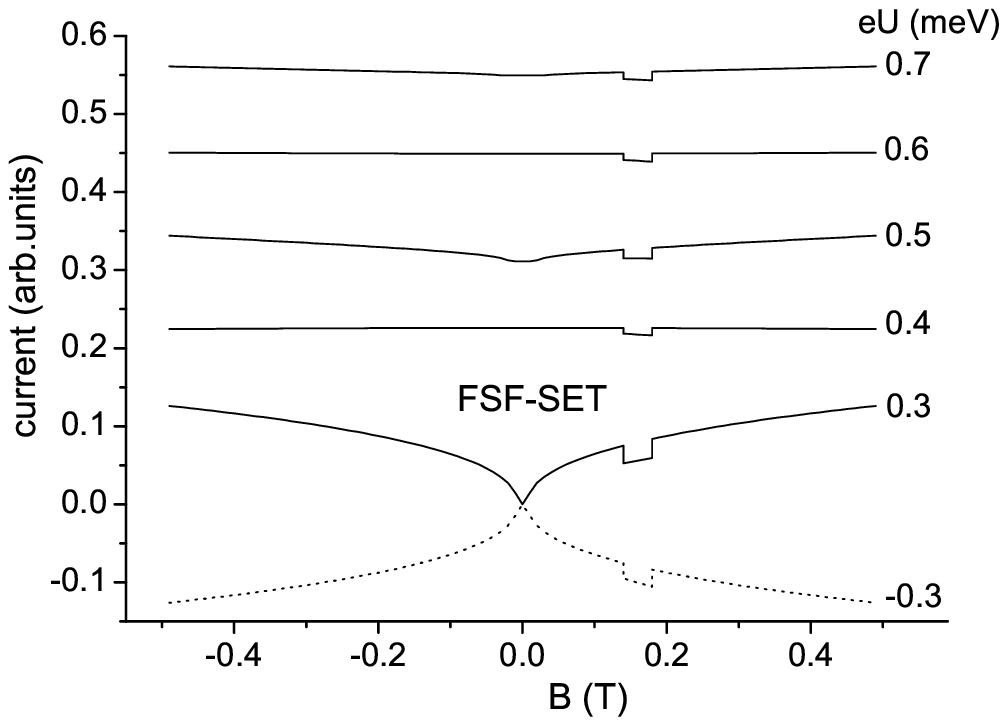}%
}%
\\
&
\begin{tabular}
[c]{l}%
Fig.6: The current through the FSF-single electron transistor\\
for different bias. The switching fields are identical to Fig.5,\\
field sweep is from $+1.5T$ to $-1.5T$. Energy gap and Coulomb\\
energy are 0.2meV and 0.1meV.
\end{tabular}
\end{align*}

There are a few kinks in the current curves of Fig.6 as a function of the
magnetic field. They occur when when the maximum number of electrons on the
island changes by one. The magnetic field lowers one subband of the
superconductor and reduces the energy gap. Whenever $\left[  \left(
eU+\mu_{B}B\right)  -\Delta\left(  B\right)  \right]  /E_{Cb}$ crosses an odd
integer $\left(  2n+1\right)  $ as a function of increasing $\left\vert
B\right\vert $ the maximum number of electrons on the island increases by one.
Furthermore one observes that again for negative bias the sign of the relative
current jump in the window $\left(  B_{sw},B_{sn}\right)  $ changes sign.

\section{Discussion and Conclusion}

In the discussion of a single ferromagnet-superconductor junction we arrived
at the following conclusions.

\begin{itemize}
\item For electron flow from the ferromagnet F into the superconductor S the
current \textbf{increases} when the magnetization\ $\mathbf{m}$ aligns
parallel to the magnetic field.

\item For electron flow from the superconductor S into the ferromagnet F the
current \textbf{decreases} when the magnetization\ $\mathbf{m}$ aligns
parallel to the magnetic field.
\end{itemize}

From these facts it follows that the current jump in a double junction changes
sign when one reverses the bias. When the source electrode (the electrode from
which the electrons tunnel into the island) flips its magnetization first then
the conductance of the source-island junction increases and therefore the
current through the SET increases. When the drain electrode (the electrode
into which the electrons tunnel from the island) flips its magnetization first
then the conductance of the island-drain junction decreases and therefore the
current through the SET decreases. Since a reversal of the bias exchanges
source and drain one finds that the relative change of the current at the
field $B_{sw}$ has the opposite sign.

In a nut shell: a flip of the magnetization in the electron source yields an
increase of the current and a flip of the magnetization in the electron drain
a decrease. The $I-B$-curves for opposite directions of the magnetic field
sweep are mirror images of each other.

In this paper we intentionally excluded a spin-coupling between the two
tunneling junctions. Such a coupling has been observed for example in the
beautifull spin precession experiment by Jedema et al. \cite{J10}. The
Meservey-Tedrow effect is an additional phenomenon which has to be included in
the analysis of FSF-single electron transistors. It can be distinguished from
the effect of spin-accumulation and gap reduction because it changes sign when
the bias voltage is reversed.

Abbreviations: SET=single electron transistor, FSF=ferromaget-superconductor-ferromagnet.

Acknowledgement: The research was supported by the National Science Foundation
NIRT program, DMR-0334231.%

\[
\]

\end{document}